\documentclass[10pt]{iopart}
\pdfoutput=1
 \usepackage{graphicx}
\usepackage{amssymb}
\usepackage{amsfonts}
\usepackage{color}
\usepackage[latin1]{inputenc}
\usepackage{float}
\usepackage{comment}

\renewcommand{\vec}{\mathbf}

\begin{document}
\title[Multiscale model for the effects of adaptive immunity on virotherapy]
{Multiscale model for the effects of adaptive immunity suppression on the viral
therapy of cancer}
\author{$^{1,2}$Leticia R Paiva, 
$^1$Hallan S Silva,
$^1$Silvio C Ferreira and
$^{1,2}$Marcelo L Martins}
\address{$^1$Departamento de F\'{\i}sica, Universidade Federal de Vi\c{c}osa, 
36570-000, Vi\c{c}osa, MG, Brazil \\
$^2$National Institute of Science and Technology for Complex Systems, Brazil}

\begin{abstract}
Oncolytic virotherapy - the use of viruses that specifically kill tumor cells -
is an innovative and highly promising route for treating cancer. However, its
therapeutic outcomes are mainly impaired by the host immune response to the
viral infection. In the present work, we propose a multiscale mathematical model
to study how the immune response interferes with the viral oncolytic activity.
The model assumes that cytotoxic T cells can induce apoptosis in infected cancer
cells and that free viruses can be inactivated by neutralizing antibodies or
cleared at a constant rate by the innate immune response. Our simulations
suggest that reprogramming the immune microenvironment in tumors could
substantially enhance the oncolytic virotherapy in immune-competent hosts.
Viable routes to such reprogramming are either in situ virus-mediated impairing
of CD$8^+$ T cells motility or blockade of B and T lymphocytes recruitment. Our
theoretical results can shed light on the design of viral vectors or new
protocols with neat potential impacts on the clinical practice.
\end{abstract}

%\submitto{\it J. Stat. Mech.: Theor. Exp.}
\section{Introduction}
Recently, cancer replaced heart disease as the leading cause of death among the
United States citizens younger than 85 years \cite{Siegel} and will probably
become the leading one in some other parts of the world within a few years
\cite{Murray}. In the past decades we have witnessed an extraordinary progress
on imaging, diagnosis, and research in the molecular biology of cancer, but its
medical treatment, especially of tumors at unresectable locations, metastasis or
recurrent neoplasias, still has many limitations \cite{Weinberg}. Even the
advent of molecularly targeted therapies has led to modest improvements for the
majority of patients with advanced cancers \cite{Chen}. Due to the complexity of
tumor growth pathways, increasing resistance and tumor progression still is the
rule for patients with metastatic disease. Hence, new pathway-independent
therapeutic agents represent a central alternative. Among them, oncolytic
viruses are unique since they can be amplified by infected cells, armed to
selectively infect and kill cancer cells and induce an immune response against
the tumor \cite{Harrington}.

Several oncolytic viruses exhibiting safety and selective replication in tumors
are under current evaluation in pre-clinical and clinical trails
\cite{Vidal,Parato}. However, durable objective responses are relatively rare
because their \textit{in vivo} antitumor efficiencies are mainly impaired by the
host immune response \cite{Hecht} that prevents successful virus spreading and
tumor remission \cite{Friedman}. In a clinical context, the therapeutic outcome
is a drastically reduced direct oncolytic activity, much less than those
suggested by experimental models. Initially, upon virus administration, the
innate immunity plays a significant role in limiting virotherapeutic efficiency
\cite{Alvarez,Nguyen}. Indeed, in gliomas treated with herpes simplex viruses
(HSV), the recruitment of infiltrating monocytic cells has been associated with
intratumoral clearance of over $80\%$ of HSV-derived viral particles shortly
after delivery \cite{Fulci}. Subsequently, the adaptive immune response develops
and becomes dominant after $\sim 5$ days post-infection \cite{Bessis}. This
cellular and humoral immune response involves the recruitment of
antigen-specific B cells and $CD8+$ T cells to the infected tissue.

As can be noticed, various fundamental issues and technical hurdles must be
understood and overcome in order to enhance the efficacy of oncolytic
virotherapy. It is imperative to enlarge our current understanding concerning
the complex and coupled dynamics of the growing tumor, the host immune response
and the oncolytic viruses. The nonlinearities and complexities inherent to such
dynamics call for mathematical approaches \cite{Martins1,Martins2}. Quantitative
models can reveal the major parameters affecting therapeutic outcomes, guide new
essays by indicating relevant physiological processes for further investigation,
and prevent excessive experimentation needed to develop effective treatments.

In the present paper, we report on theoretical findings derived from a
multiscale agent-based model for oncolytic virotherapy. Such findings point out
implications for the design of new replication-competent viruses or combined
therapy aimed to modulate the immune response. In section \ref{model}, the model
is described. It takes explicitly into account the individual, discrete nature
of the oncolytic viruses and includes the host adaptive immune response against
the viruses. In section \ref{immunity}, the simulation results are discussed. In
section \ref{clinic}, current or viable strategies to enhance the virotherapy
efficacy against solid tumors consistent with our simulation results are
emphasized. Finally, some conclusions are drawn in section \ref{summary}.

\section{A multiscale agent-based model for oncolytic virotherapy}
\label{model}

The proposed model is an extension of our work on oncolytic virotherapy
\cite{Paiva,Paiva2,Ferreira} and involves multiple agents and processes involved
at distinct time and length scales.

\subsection{The tissue}
The tissue is modeled by a square lattice fed through a single capillary vessel
at the top of the lattice. Six different cell types (normal and dead cells,
uninfected and infected cancer cells, B and $CD8+$ T lymphocytes) and an
oncolytic virus are considered. Each cell and virus are represented by
individual agents and their populations in a site $\vec{x}=(i,j)$ described by
$\sigma_n$, $\sigma_d$, $\sigma_c^{un}$, $\sigma_c^{inf}$, $\sigma_B$,
$\sigma_{T}$, and $\sigma_v$, respectively. Only one cell can occupy a given
site, except uninfected or infected cancer cells that can pile up because tumor
cell's division is not constrained by contact inhibition. Antibody molecules
directed against the oncolytic virus are also considered and described by its
concentration $A(\vec{x})$. Since the viruses and antibodies are very small
particles in comparison with cells, there is no constraint on their populations
or concentrations at any site.

The nutrients, diffusing from the capillary vessel throughout the tissue, are
divided into two groups: those essential to maintain the basic cell functions,
whose deprivation can induce death ($\phi_1$) and those that limit cell
replication but are not demanded for cell survival ($\phi_2$). Both nutrient
types are described by continuous fields $\phi_j(\vec{x}, t)$, which evolve in
space and time accordingly the simplest (linear with constant coefficients)
dimensionless reaction-diffusion equations

\begin{equation}
\frac{\partial \phi_j}{\partial t}= \nabla^2 \phi_j - \alpha^2 \phi_j \sigma_n -
\lambda_j \alpha^2 \phi_j \sigma_c,
\label{dif_nut}
\end{equation}
with distinct uptake rates for normal and cancer cells and a characteristic
length scale $\alpha$ for nutrient diffusion \cite{Ferreira2}. The local
populations of normal and cancer cells are $\sigma_n=0$ or $1$ and
$\sigma_c=\sigma_c^{un}+\sigma_c^{inf}=0,1,2,\ldots$, respectively. Equation
(\ref{dif_nut}) obeys a periodic boundary condition along the direction parallel
to the capillary and a Neumann boundary condition at the border of the tissue
($i=L$). At the capillary ($i=0$), the concentration is $\phi_{1,2}=1$
(continuous and fixed supply).

\subsection{Cancer growth}
The tumor grows from a single malignant cell according to a stochastic dynamics
whose probabilities depend on the local nutrient concentration. Each uninfected
cancer cell, randomly selected with equal chance, can carry out one of four
actions:

\noindent 1- \textbf{Mitotic replication}, with a probability

\begin{equation}
P_{div} =1-\exp \left[-\left( \frac{\phi_2}{\theta_{div}\sigma_c} \right)^2
\right],
\label{pdiv}
\end{equation}
an increasing function of the nutrient concentration per cancer cell, $\phi_2$.
The daughter cell randomly occupies one of its normal or necrotic nearest
neighbor sites, where there exists any, or otherwise piles up at its mother
site.

\noindent 2- \textbf{Death}, with a probability

\begin{equation}
P_{del} =\exp \left[-\left( \frac{\phi_1}{\theta_{del}\sigma_c} \right)^2
\right],
\label{pdel}
\end{equation}
that increases with the scarcity of nutrients $\phi_1$ essential to sustain the
cell metabolism.

\noindent 3- \textbf{Migration}, with a probability

\begin{equation}
P_{mov} =1- \exp \left[-\sigma_c \left( \frac{\phi_1}{\theta_{mov}} \right)^2
\right],
\label{pmov}
\end{equation}
that increases with the local population of cancer cells and the nutrient
concentration per cell. The migrating cell moves to one of its nearest-neighbor
sites chosen at random, interchanging its position with a normal or dead cell if
there exists any. The interchanged normal cell is eliminated when it arives at a
site still occupied by other cancer cells. A probability increasing
with the nutrient concentration is justified by the necessity
of nutrients for cell motility and, in addition, by the degradation 
of the extracellular matrix near the tumor surface that
releases several chemicals that promote cell migration and
proliferation. This hypothesis is consistent with experimental
data in multicellular tumor spheroids~\cite{Freyer} and was previously
used in other mathematical models~\cite{Kansal,Ramis}.

\subsection{Oncolytic virotherapy}
The virotherapy begins when the tumor attains $N_0$ cells and consists in a
single virus injection. In the direct intratumoral administration, viruses are
uniformly spread over the entire tumor at a given multiplicity of infection
(MOI). Thus, $N_0 \times MOI$ viruses are initially released. This approach
corresponds to the experimental protocols used in severe combined immune
deficient (SCID) mice \cite{Coffey} and in vitro assays \cite{Bischoff,Raj}.

After virus administration, each uninfected cancer cell, randomly selected with
equal chance, can \textbf{become infected} with a probability
\begin{equation}
P_{inf} =1- \exp \left[-\left( \frac{\sigma_v}{\sigma_c \theta_{inf}} \right)^2
\right],
\label{pinf}
\end{equation}
an increasing function of the local viral load per cell, controlled by the
parameter $\theta_{inf}$. The model assumes perfect viral selectivity for cancer
cells, thereby the infection of normal cells by oncolytic viruses is neglected.
The number of viruses $n_v$ that infect a given cell is selected from a Poisson
distribution,
\begin{equation}
P(n_v) = \frac{k^{n_v} e^{-k}}{n_v!},
\end{equation}
where $k$ is the typical viral entry \cite{Dixit}. The model assumes that an
infected cancer cell neither divides nor migrates because its slaved cellular
machinery is focused on virus replication. It is also assumed that infected
cancer cells sustain their metabolism until lysis and die only by lysis. The
\textbf{death by lysis} occurs with a probability
\begin{equation}
P_{lysis} = 1- \exp \left( -\frac{T_{inf}}{T_l} \right),
\label{plysis}
\end{equation}
where $T_{inf}$ is the time elapsed since the cell infection and $T_l$ is the
characteristic time for cell lysis. The lysis of each infected cancer cell
releases
\begin{equation}
v_0 = b_s \frac{n_v}{\xi + n_v}
\end{equation}
free viruses to the extra-cellular medium. Here, the maximum virus burst size
$b_s$ and $\xi$ are model parameters. At the time of lysis, the new free viruses
remain on the site of the lysed cell. At each time step, these free viruses
either spread through the tissue by performing independent random walks
comprising $q$ steps or are cleared at a rate $\gamma_v$. The clearance rate
$\gamma_v$ embodies the innate immune responses and non immune mechanisms of
virus inactivation.

\subsection{Antiviral immune response}
Concerning the adaptive immune response, the model assumes that B and CD$8^+$ T
cells are recruited to the infected tissue if there are more than
$N_{inf}^{min}$ infected cancer cells within the tumor tissue. At lower
infection levels, the signaling is supposed too weak to induce lymphocyte
recruitment, and the populations $N_B$ of B and $N_T$ of CD$8^+$ T lymphocytes
will decrease due to their finite lifespans, $T_B$ and $T_{T}$, respectively.
The recruited lymphocytes extravagant from the capillary vessel into the tissue
at constant rates $\beta_B$ and $\beta_T$, respectively. This is the simplest
description of the complex process of lymphocyte recruitment.

At each time step, $N_{T}$ effector $CD8+$ T cells randomly selected move inside
surveillance areas of radii $R_{T}$ centered on their present position. In every
area free from infected cancer cells, the chosen T cell jumps at random to a
site at its border. Otherwise, the cytotoxic T cell jumps randomly to an empty
nearest-neighbor site of an infected cancer cell inside that region. This
corresponds to a directed migration towards the infected region, with either
less or more confined trajectories in regions free or containing infected cancer
cells, respectively. These rules are suggested by \textit{in vivo} imaging of
cytotoxic T cell infiltration in a solid tumor growing subcutaneously
\cite{Boissonnas}. CD$8^+$ T cells attached to one or more nearest-neighbor,
infected cells will induce the apoptosis of only one of these target cells with
a probability $P_{apop}$. Indeed, the action of a cytotoxic CD$8^+$ T cell seems
to be narrowly focused on just one of its points of contact with a target cell
at once \cite{Janeway}. Again, the target cell to be destroyed will be chosen at
random among these nearest neighbors.

Similarly, every B cell can only move inside a surveillance area of radius $R_B$
either towards a site containing infected cancer cells, or to a border site
randomly chosen in infected cell free regions, where it secrets antibodies at a
rate $\beta_{A}$. In addition, a significant amount of these antibodies enter
into the systemic circulation and during the infection sustain a specific
antibody concentration in the capillary vessel. The model assumes that these
antibodies infiltrate the infected tissue at a constant rate. Thus, the antibody
concentration is given by a non-stationary solution of the diffusion equation
\begin{equation}
\frac{\partial A}{\partial t}=D_{A}\nabla^2 A +\beta_A \sigma_B -\gamma_A A .
\label{ea}
\end{equation}
Here, $D_A$ is the diffusivity and $\gamma_A$ the clearance rate of the
antibodies. At the capillary, the boundary condition is $A(x,0)=A_0$,
representing a fixed concentration in the bloodstream.

The antibodies prevent virus binding to cellular receptors and subsequent cell
entry, neutralizing the virus. We assume that free viruses can be inactivated by
antibodies with a probability
\begin{eqnarray}
P_{intact}=\left\lbrace \begin{array}{ll}
\frac{2A(i,j)}{A(i,j)+A_{inact}} &  \mbox{, if  }  A(i,j) \le A_{inact}\\
1  & \mbox{, otherwise}\\
\end{array}
\right.
\label{pinat}
\end{eqnarray}
a function of the local antibody concentration that increases monotonicaly up to
a threshold $A_{inat}$ and saturates at the unity above this concentration.
Antibodies are removed at binding to a free virus in order to inactivate it. We
assume that the local concentration of antibodies decreases by $A_{inact}$ at
each virus neutralizing event. Finally, the parameter estimates can be found in
the appendix.

\subsection{Simulation protocol}
The simulations were implemented as follows. At each time step, Eq.
(\ref{dif_nut}) is numerically solved in the stationary state ($\partial
\phi/\partial t = 0$) through relaxation methods, and Eq. (\ref{equation_a}) is
iterated for a defined number of steps. Then, fractions of free viruses are
cleared, innactivated by antibodies with a probability given by eq.
(\ref{pinat}), and each remaining virus performs a random walk with $q$ steps.
Provided the nutrient concentration and viral load at any lattice site, both B
and $CD9+$ T lymphocytes are recruited to the tissue. So, $N_1$ cancer cells and
$N_2$ lymphocytes are sequentially selected at random with equal probability.
(Here, $N_1$ is the total number of tumor cells, uninfected or infected, and
$N_2$ is the number of B and $CD8+$ T lymphocytes at the time $t$.) For each one
of them, a tentative action is randomly chosen with equal probability. These
actions are division, death, migration or infection for an uninfected cancer
cell, lysis for an infected tumor cell, movement for a B cell and movement or
cytotoxic action for a $CD8+$ T lymphocyte. For a cancer cell, the selected
action will be implemented or not according to the corresponding local
probabilities determined by Eqs. (\ref{pdiv}) to (\ref{plysis}). In the case of
infection, an integer random number $n_v$ Poissonian distributed is generated
and compared with the local virus population $\sigma_v$. If $\sigma_v \ge n_v$,
$n_v$ viruses invade the selected uninfected cancer cell, decreasing $\sigma_v$
by $n_v$. Otherwise, this process will be repeated until generates a $n_v \le
\sigma_v$. In turn, in the case of lysis, $v_0$ new viruses are introduced at
the site of the lysed cell. For a T cell cytotoxic action, the apoptosis of a
nearest-neighbor infected cancer cell is elicited with a probability $P_{apop}$.
 At the end of this sequence of $N_1+N_2$ updates, a new time step starts and
the entire procedure is iterated. The simulations stop if any tumor cell reaches
the capillary vessel or the tissue border or if the tumor is eradicated
($N_c=0$).

\section{Hurdling the major obstacle: immunity}
\label{immunity}
We simulated the model considering a viral agent able to eradicate compact solid
tumors in the absence of an adaptive immune response, as reported in reference
\cite{Paiva}. When applied to immune-competent hosts characterized by typical
immunological parameters (see the Supplementary Information), a virotherapy
based on such oncolytic virus invariably fails. The reason is that free viruses
as well as their sources, the infected cancer cells, are eliminated by the
adaptive immune response. Even at small virus clearance rate, representing a
permanent depletion of the innate response, the adaptive immunity against the
oncolytic virus ends to subvert the therapy. As experimentally observed
\cite{Fulci}, there occurs only a significant transient increase in intratumoral
viral titers shortly after viral administration. Hence, the central issue is how
to suppress, or better, modulate the adaptive immune response in order to
enhance the success of viral therapies of cancer in immuno-competent hosts.

\begin{figure}
\begin{center}
\includegraphics[width=10cm]{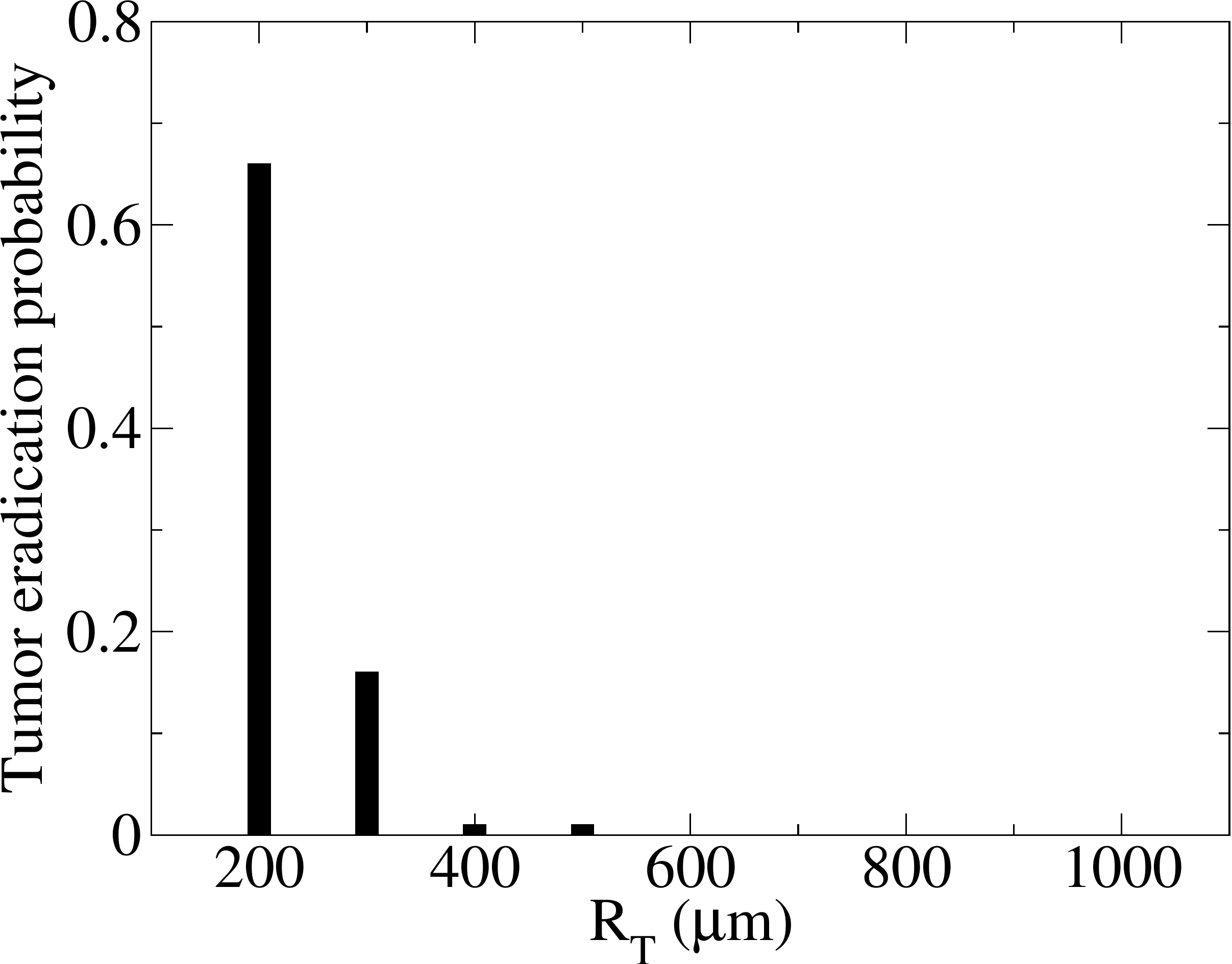}
\caption{\label{noBcell}Tumor eradication probability as a function of the
CD$8^+$ surveillance radius. The recruitment of B cells is suppressed
($\beta_B=0$) and that of T cells occurs at a rate $\beta_{T}=40$ ($\equiv
10h^{-1}$). Antibodies infiltrate the tissue from the capillary, where their
concentration is $A_0$. The data correspond to $100$ independent simulated
samples.}
\end{center}
\end{figure}

Our results indicate three routes to enhance the chances of tumor eradication in
immuno-competent patients. In the first one, the recruitment of B cells to the
infected tissue is completely suppressed. The corresponding probabilities of
therapeutic success for different CD$8^+$ T cell surveillance radii are shown in
figure \ref{noBcell}. The absence of B cells ensures that the antibody
concentration within the tissue remains always below the level necessary to
neutralize the free oncolytic viruses, whereas short-range surveillance radii
make certain that CD$8^+$ T cells reach the tumor at a rate insufficient to
eliminate the infected cancer cells before the successive waves of viruses
infect all cancer cells. In contrast, cytotoxic T cells having a long-range
surveillance radius ($R_{T}\geq300\mu m$) can induce apoptosis in all infected
tumor cells if recruited at rates $\geq 5$ cells$/h$. Now, the successive rounds
of infection are strongly attenuated and the probability of therapeutic success
becomes lower than $15\%$. Furthermore, the oncolytic virotherapy fails if the
concentration of antibodies at the capillary is very large, as possibly elicited
by a potent systemic immune response. Indeed, the probability of therapeutic
success decreases to about $10 \%$ if $A_0$ is tenfold greater and vanishes if
$A_0$ is multiplied by a factor $\sim 10^3$, a concentration corresponding to
the levels observed in a secondary immune response to a previously presented
antigen. These results are consistent with clinical trials and rodent tumor
models, which indicate that neutralizing antibodies against viral antigens
significantly impair the therapeutic efficacy of several adenoviral vectors
\cite{Vidal,Bessis,Alemany}.

\begin{figure}
\begin{center}
\includegraphics[width=10cm]{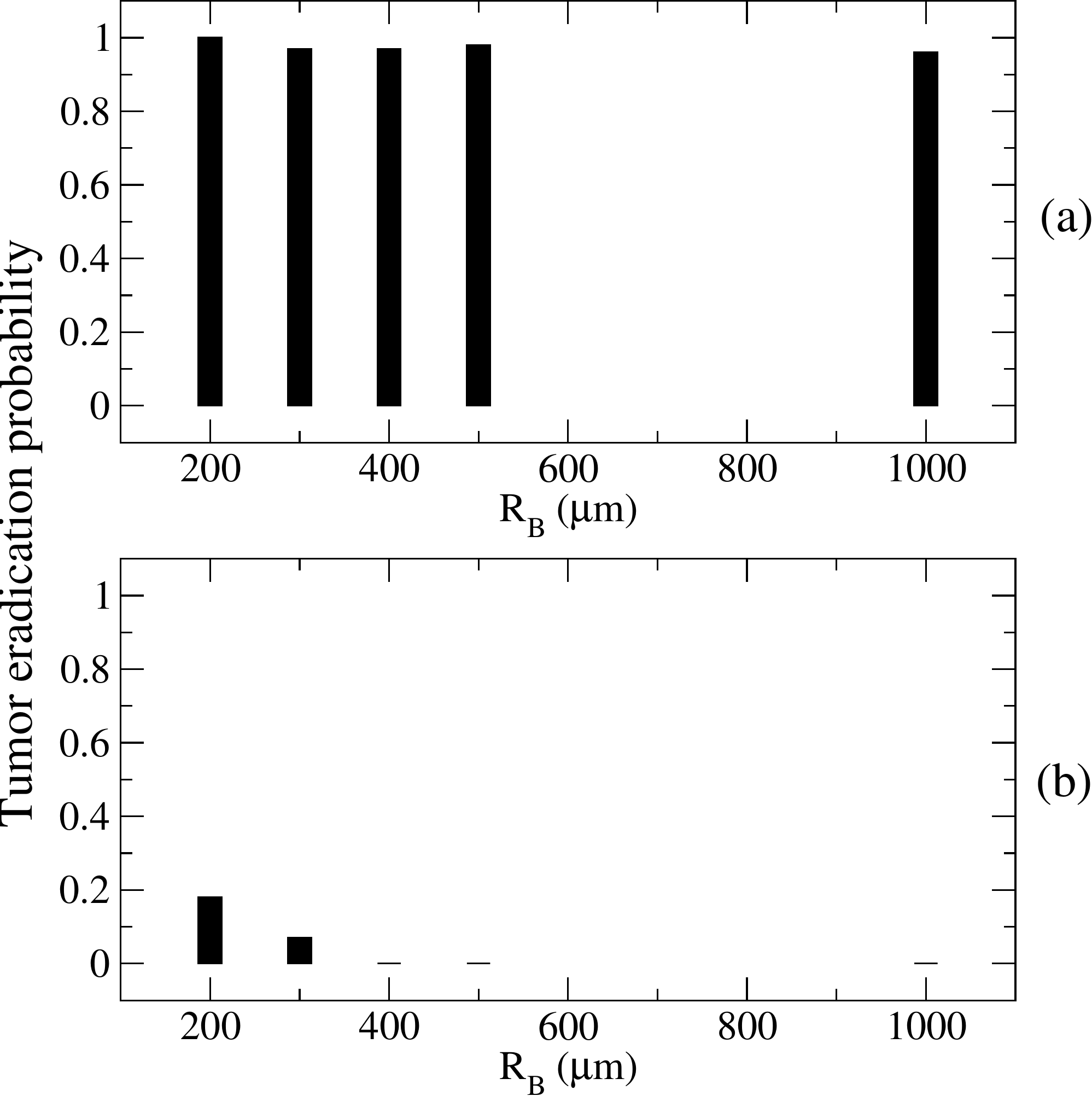}
\caption{\label{noTcell}Tumor eradication probability for increasing B cell
surveillance radius. The recruitment of T lymphocytes is suppressed
($\beta_T=0$) and antibodies (a) do not infiltrate the tissue or (b) extravasate
from the capillary into the tissue. The antibody's concentration at the
capillary $A_0$. B cells are recruited at a rate $\beta_{B}=10h^{-1}$. The data
correspond to $100$ independent simulated samples.}
\end{center}
\end{figure}

The second route consists in suppressing both CD$8^+$ T cells and antibody
infiltration. Thus, only B cells are recruited to the tumor site. In
consequence, the tumor is eradicated  with probability higher than $90\%$ (see
Figure \ref{noTcell}a). This great therapeutic success is sustained even for B
cells with surveillance radius as large as $R_B= 500 \mu m$, recruited at rates
up to $\beta_B=40h^{-1}$, and synthesizing antibodies at rates up to two orders
of magnitude greater than $\beta_A$. Our simulations indicate that B lymphocytes
alone can not elicit an effective immune response against the oncolytic virus.
Indeed, B lymphocytes are antibody's sources in permanent migration. They do not
stay in the neighborhood of any infected cancer cell long enough to increase
sufficiently the local antibody's concentration to inactivate most of viruses
released after the lysis of this cell. However, if antibodies extravasate from
the capillary into the tissue, the total antibody concentration in the tumor is
high enough to inactivate most of the free viruses, decreasing abruptly the
therapeutic success (figure \ref{noTcell}b).

\begin{figure}
\begin{center}
\includegraphics[width=10cm]{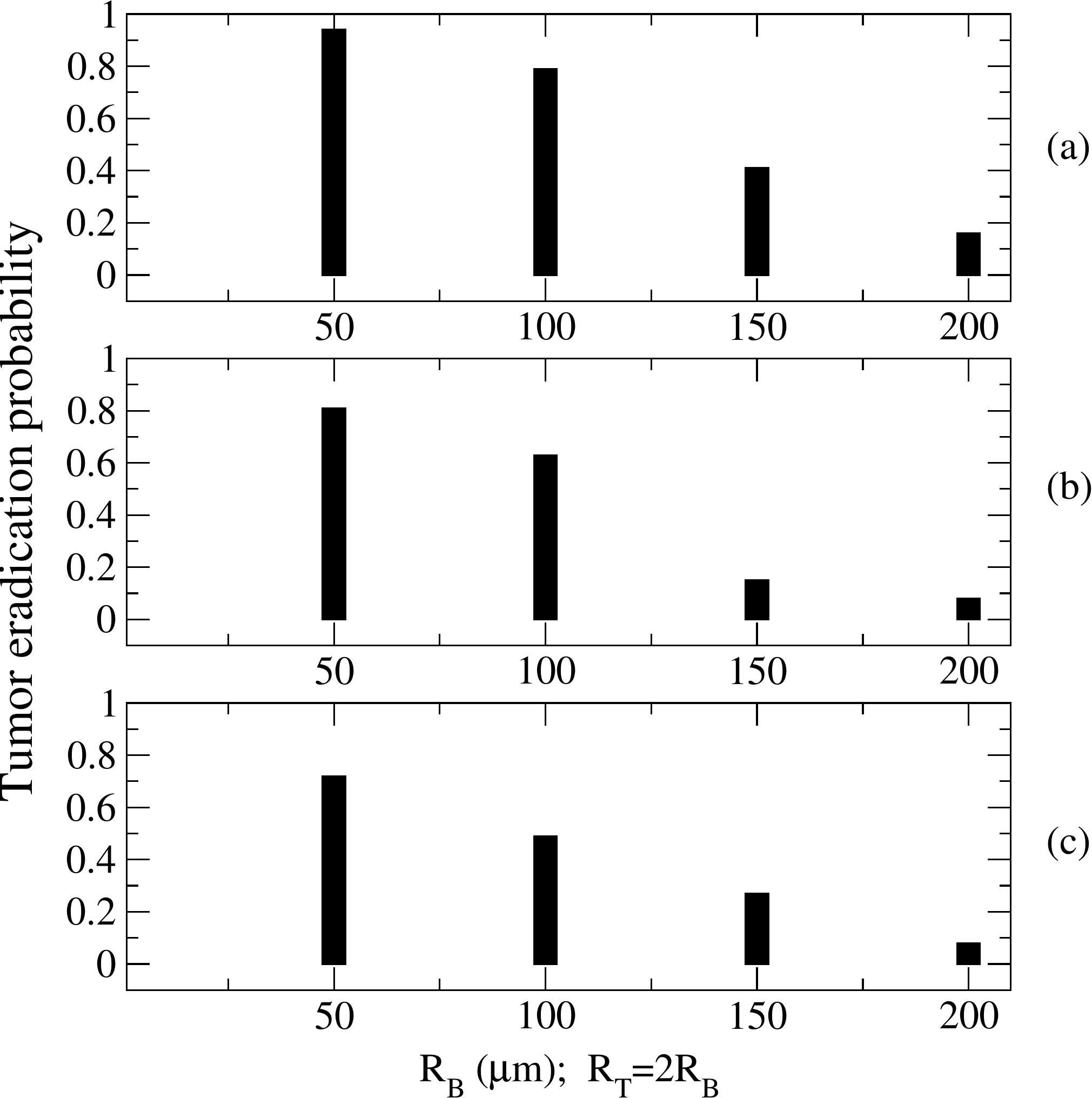}
\caption{\label{BTmot}Tumor eradication probability as a function of lymphocyte
surveillance radii. Antibodies infiltrate the tissue from the capillary, where
$A=A_0$. The lymphocytes are recruited at rates (a)
$\beta_{B}=\beta_{T}=1.25h^{-1}$; (b) $\beta_{B}=\beta_{T}=2.5h^{-1}$; and (c)
$\beta_{B}=\beta_{T}=5h^{-1}$. The data correspond to $100$ independent
simulated samples.}
\end{center}
\end{figure}

\begin{figure}
\begin{center}
\includegraphics[width=13cm]{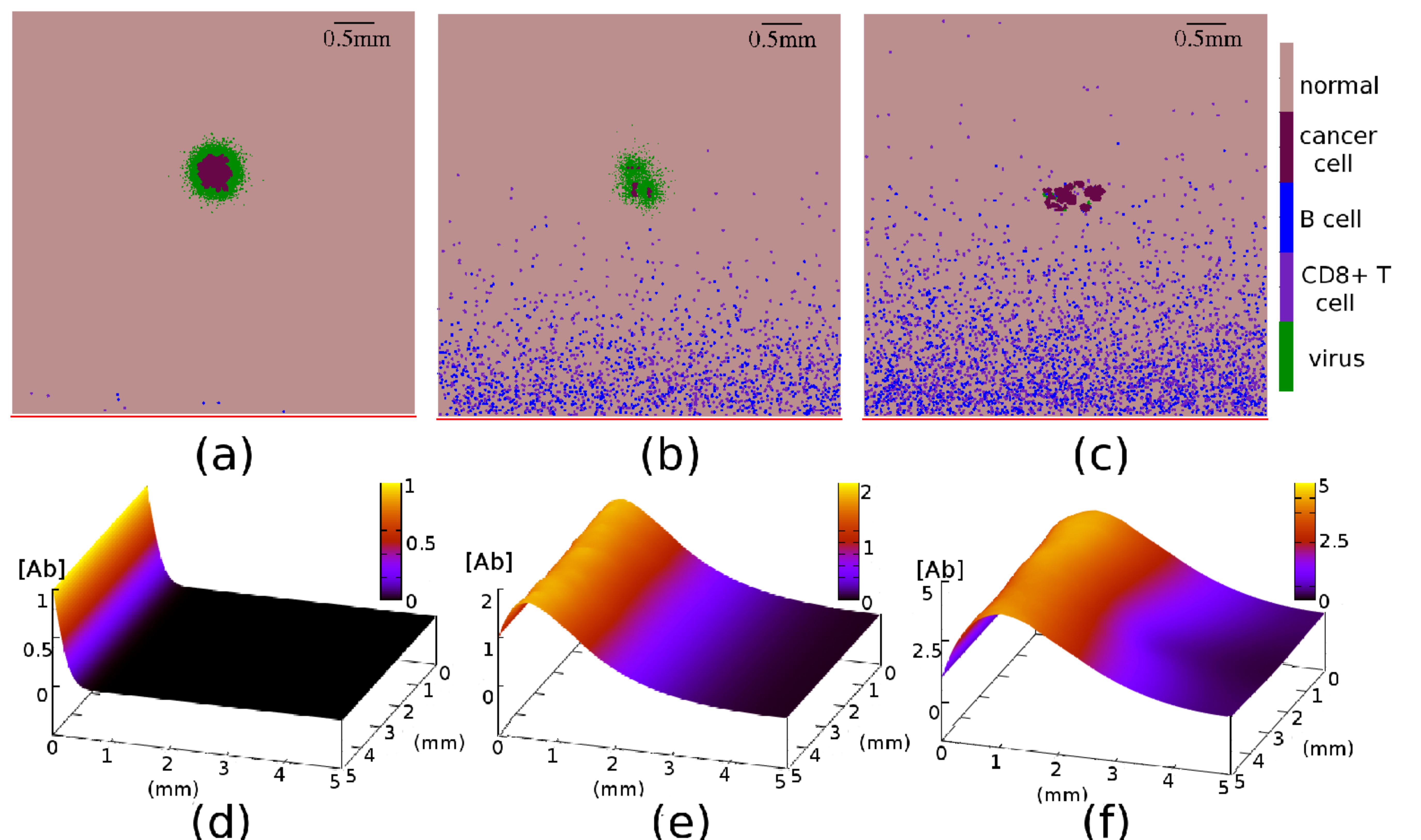}
\caption{\label{BTpatterns}Spatiotemporal patterns of cancer cells, lymphocytes
and virus ((a),(b) and (c)) and related antibody concentrations ((d),(e) and
(f)). (a) and (d): the beginning of the adaptive immune response, $5$ days after
the virus administration; (b) and (e): $20$ days later. At this time the tumor
is almost eradicated. (c) and (f): $2$ months later. The adaptive immune
response eliminated almost all free viruses and infected cancer cells, but the
tumor keeps growing. The values $\beta_B=\beta_{T}=2.5h^{-1}$ for lymphocyte
recruitment rates, $R_{B}=150 \mu m$ and $R_{T}=300 \mu m$ for B and T cells
surveillance radius were used. For this parameter set, the probability of
therapeutic success is $30\%$. The red line in top panels represents the
capillary vessel.}
\end{center}
\end{figure}

Finally, the third scenario for tumor eradication consists in impairing the
motility of both B and T cells without affect the antibody dynamics or 
lymphocyte recruitment rates. If the surveillance radii are short-ranged 
($R_B\leq 100 \mu m$ and $R_{T}\leq 200 \mu m$) the tumor is eradicated with a
probability greater than $70\%$ for lymphocytes recruited at low rates
$\beta_B=\beta_{T}=5h^{-1}$. In figure \ref{BTmot}, the tumor eradication
probability is shown for three distinct recruitment rates. Lymphocytic motility
is a much more determinant feature than its recruitment rate. The key to success
is to provide enough time for the viruses to kill the cancer cells before the
infiltrating B and T lymphocytes elicit an effective antiviral cellular immune
response. If a few cancer cells survive the virus attack until they fall into
the surveillance radii of the incoming B and T cells, the virotherapy fails
because the free viruses and infected cancer cells are then quickly eliminated.
This is illustrated in figure \ref{BTpatterns}. As can be noticed, $20$ days
after viral administration the tumor undergoes a significant remission and the
first lymphocytes have the tumor inside their surveillance radii. Then, the
local adaptive immune response mediated by B and T cells spends around $2$
months to eliminate almost all free viruses and infected cancer cells, leaving
room for tumor regrowth from the few surviving malignant cells.

\section{From theory to the clinical practice}
\label{clinic}
Numerous clinical trials have firmly demonstrated that the efficacy of oncolytic
viruses is drastically reduced by the host antiviral immune response.
Accordingly, the immunossupressor cyclophosphamide, a chemotherapeutic drug
which inhibits the synthesis of neutralizing antibodies \cite{Hirasawa,Jooss},
IFN-$\gamma$ \cite{Ikeda2}, and the activity of innate immune cells
\cite{Fulci2}, is commonly co-administered with HSV \cite{Ikeda,Wakimoto} and
reovirus \cite{Parato}, abrogating the innate and adaptive humoral immunity.
However, instead of a systemic immunity suppression, our results support that a
local, \textit{in situ}, disruption of the antiviral response strongly enhances
oncolytic virotherapy.

This goal can be achieved through the design of efficient oncolytic viruses. 
Indeed, one rationale supported by our simulational results is to arm oncolytic
viruses with genes encoding for inhibitors of chemokines, interleukines, or
interferons. Once introduced in, expressed and secreted by infected cancer cells
up to their lysis, such inhibitors will actively abrogate the recruitment of
lymphocytes, mainly effector CD$8^+$ T cells, to the tumor site. This goal can
possible be achieved through diverse approaches. For instance, by blocking the
expression of CXCL9 and CXCL10 \textit{in situ}. These chemokines are necessary
for cytotoxic T lymphocytes (CTL) to enter from the systemic circulation into
the site of infection \cite{Nakanishi}. Moreover, the down-regulation of CXCL9
reduces CD$4^+$ T cell infiltration into the infected tissue \cite{Wuest}. Such
local depletion of CD$4^+$ T cells significantly blocks further CTL recruitment,
primarily dependent on the IFN-$\gamma$ secreted by the CD$4^+$ cells
\cite{Nakanishi}. In turn, lower local levels of CD$4^+$-derived IFN-$\gamma$
inhibits the expression of CXCL9 and the autocrine-mediated conditioning of
CD$4^+$ T cells for access to the infected tissue \cite{Nakanishi}, closing a
feedback loop. Even the residual CTL recruitment observed in the absence of
CD$4^+$ lymphocytes or IFN-$\gamma$ can be eliminated through the neutralization
of type I IFN, such as IFN-$\alpha \beta$R 
\cite{Nakanishi}. 

Related to this context, one innovative development deserve special emphasis.
the rVsV-gG virus inhibited the chemotaxis of NK and NKT cells to the tumor
sites \cite{Altomonte}. This gG viral protein binds C, CC and CXC chemokines
with high affinity, possibly suppressing lymphocytic recruitment in addition to
the host inflammatory response. Such a possibility remains to be checked.
Concerning interleukines, several groups \cite{Ziauddin} have reported that
vaccinia virus expressing Th2 cytokines such as IL-4 and IL-10 has increased
\textit{in vivo} viral replication and slowed clearance rate. Maybe, another
inspiring case is that of a HSV-1 vector expressing IL-5 able to decrease the
numbers of infiltrating lymphocytes in the brain during the treatment of
experimental autoimmune encephalomyelitis \cite{Nygardas}.

The second rationale lies on severely constrain the motility of B and T cells
that infiltrate the infected tissue. A possible approach is engineering
oncolytic viruses to encode potent and diffusible products disruptive of
migration pathways targeted to the effector lymphocytes. At the site of
infection, the lymphocytes' adhesion molecules, particularly $\beta2$-integrins,
turn into highly active states. Again, this activation is induced by chemokines
\cite{Woolf} and Rap1 is a critical mediator of both $\beta1$- and
$\beta2$-integrin activation, cell polarization  as well as cell migration
\cite{Kinashi}. Thereby, presumably an oncolytic vector encoding an inhibitor of
Rap1 or its effector molecule RAPL can significantly impair lymphocyte motility.
Alternative strategies might target the lymphocyte's cytoskeleton. T cells use
amoeboid migration based on actin polymerization along the plasma membrane to
stiffen and contract the cell cortex \cite{Friedl}. Small GTPase RhoA and its
effector ROCK critically control the cortical actin dynamics. Interfering with
RhoA and/or ROCK results in a failure of the trailing edge to detach,
dramatically reducing the migration rate of T cells \cite{Smith}, as well as
impairing acto-myosin contraction required for amoeboid movements. Thus,
inhibitors of such proteins emerge as immediate transgene candidates for
virus-mediated suppression of lymphocytic motility. Other candidate can be, for
instance, the Wiskott-Aldrich syndrome protein (WASP) that plays a pivotal role
in regulating surface receptor signaling to the actin cytoskeleton in
hematopoietic cells. Blocking WASP activity leads to a severe defect in cell
migration in multiple cell types \cite{Snapper}.

\section{Concluding remarks}
\label{summary}
A fundamental issue concerning the efficiency of the viral therapy of cancer,
namely, the abrogation of the host antiviral immune response, was investigated
through computer simulations of a multiscale model for tumor growth. The model
combines macroscopic diffusion equations for the nutrients and antibodies and
stochastic rules for the actions of individual cells and viruses. Even though
current strategies to improve virotherapeutic efficacy consist in developing
viruses that can evade the adaptive immune response, we propose testable
alternatives focused instead on the local disruption of the cellular-mediated
immunity.

Essentially, our results indicate two distinct and complementary routes. The
first one is to halt the recruitment of effector lymphocytes, mainly the CD$8^+$
T cells, at the tumor-bearing tissue infected with the oncolytic virus. The
second approach is based on strongly reducing the motility of B and T
lymphocytes that infiltrate the affected tissue. We find that the lymphocytes
surveillance radii, determining both their motility and targeting ability,
rather than their recruitment rates, are the major immunological features to be
modulated. Hence, the second route seems to be the most effective way to enhance
the success of oncolytic virotherapy. However, the potential benefit of using
these two strategies in synergy is promising.

Finally, our results fuel the debate on apparently disparate strategies to
enhance oncolytic virotherapy: develop oncolytic viruses competent to either
re-stimulate antitumoral immune response or inactivate antiviral immunity. Both
are currently under experimental test \cite{Altomonte,Zamarin,Choi}. Given the
complexity of tumor-stroma-immune system interactions, mathematical modeling can
help us to evaluate quantitatively these strategies or even the result of their
combination.

\section*{Acknowledgments}
This work was partially supported by the Brazilian Agencies CAPES, CNPq, FAPEMIG
and FACEPE.

\appendix*
\section{Parameter estimates}

The model parameters determining the cancer growth were fixed at $\alpha=1/L$,
$L=500$, $\lambda_1=25$, $\lambda_2=10$, $\theta_{div}=0.3$, $\theta_{mov}=5$,
$\theta_{del}=0.03$ in order to generate compact, solid tumors \cite{Ferreira2}.
In all the simulations, the virus parameters were fixed in $T_{l}=16$ time steps
($\sim 64$ h), $\theta_{inf}=0.01$, $MOI=3$, $k=3$, $bs=100$, $\xi=0.5$, and
$q=4$. Also, $\gamma_v=0.03$ was used, resulting in a clearance rate of
$7.5\times10^{-3}h^{-1}$, or only $30\%$ of its typical value \cite{Friedman}.
This value corresponds to a weak innate immune response. The virotherapy begins
when the tumor contains $N_0=5,000$ cancer cells.

Estimates for the values of the parameters associated to the humoral and
cellular immune response are shown in Table \ref{parameter}. They were used
either to guide the choice or to determine the range of the model parameters
used in the simulations as described next. 

\begin{table}[hbt]
\caption{Typical parameters characterizing the adaptive immune response.
Abbreviation: Ab $-$ antibody.}
\begin{center} 
\begin{tabular}{|c|c|c|c|}
\hline Parameter & Range of values & Description & Ref. \\
\hline $T_B$ & $<8$ weeks & B cells' lifespan & \cite{Janeway} \\ 
\hline $T_{T}$ & $4-8$ weeks & $CD8+$ T cells' lifespan & \cite{Janeway} \\ 
\hline $\beta_B$ & $1.25-10$ $h^{-1}$ -  & B cells recruitment rate &
\cite{Nakanishi} \\ 
\hline $\beta_{T}$ & $1.25-10$ $h^{-1}$ & $CD8+$ T cells recruitment rate &
\cite{Nakanishi} \\
\hline $\overline{v}$ & $8\pm 3 \mu m/min$ & $CD8+$ T cells speed within a tumor
& \cite{Boissonnas} \\
\hline $\beta_A$ & $6.944\times10^{-13}g/ml/s$ & Ab secretion rate &
\cite{Racanelli} \\ 
\hline $A_0$ & $2.5\times 10^{-9} g/ml$ & Ab concentration at the capillary &
\cite{Janeway} \\ 
\hline $\gamma_A$ & $2.31\times 10^{-7}s^{-1}$ & Ab clearance rate &
\cite{Figge} \\ 
\hline $D_A$ & $10^{-8}-10^{-7} cm^2/s$ & Ab diffusion constant & \cite{Berk} \\
\hline 
\end{tabular}
\end{center}
\label{parameter}
\end{table}

The antibody concentration needed to inactivate a virus was assumed to be
$A_{inact}=1.25\times10^{-6}g/ml$; the life times of the lymphocytes were fixed
in $T_B=T_{T}=200$ time steps, corresponding approximately to $5$ weeks; the
probability of $CD8+$-mediated infected cell apoptosis was $P_{apop}=90\%$; and
the minimum number of infected cells needed to activate an immune response was
assumed to be $N_{inf}^{min}=20$.

Nakanishi et al. \cite{Nakanishi} analyzed the $CD8+$ T cells entry into into
HSV-infected vaginal tissue and their results indicate that these lymphocytes
are recruited at a rate of $2.5$ cells$/h$. So, in the model the $\beta_{B}$ and
$\beta_{T}$ values are in the range between $1.25 h^{-1}$ and $10 h^{-1}$.
Considering the mean speed of $CD8+$ T cells migrating within a solid tumor, a
length step of $10\mu m$ (the lattice constant), and a simulation time step of
$\Delta t=4$h, lower and upper bounds for the $CD8+$ surveillance radius can be
estimated. These bounds are given by the mean squared distance traversed by a T
cell following a random and a ballistic trajectory, respectively. For the last
case, this distance is essentially $d=\overline{v} t= 1920 \mu$m$=192 \Delta$,
correspondent to $192$ steps and to a upper bound $R_{T}=192 \Delta$. Instead,
assuming a random walk with the same number $N$ of steps and that
$d=\sqrt{\langle x^2 \rangle} \sim \sqrt{N}$, a lower bound $R_{T} \approx 13
\Delta$ is found. Accordingly, the $CD8+$ T cells surveillance radius $R_{T}$
ranged from $10 \Delta$ to $200 \Delta$ in simulations. Also, since each T cell
has only $50\%$ of chance to migrate at each time step whereas B cells surely
move, it was assumed $R_B=R_{T}/2$  in order to both lymphocytes exhibit roughly
the same mean speed.

Finally, the diffusion equation for the antibodies, Eq. $9$, can be written in a
dimensionless form by performing the variable transformation
\begin{equation}
t'=\frac{D_A t}{\Delta^2}; x'=\frac{x}{\Delta}; A'=\frac{A}{A_0}.
\end{equation}
After some algebraic manipulation, one obtains the equation
\begin{equation}
\frac{\partial A'}{\partial t'}=\nabla^2 A' +\beta'_A \sigma_B -\gamma'_A A' ,
\label{equation_a}
\end{equation}
subjected to the boundary condition at the capillary vessel
\begin{equation}
A'(x,0)=1, \nonumber
\end{equation}
and involving the dimensionless rates of antibodies synthesis and clearance
\begin{equation}
\beta'_A=\frac{\Delta^2 \beta_A}{D_A A_0}; \gamma'_A=\frac{\Delta^2
\gamma_A}{D_A}.
\end{equation}
Using the experimental values shown in Table \ref{parameter}, such dimensionless
rates were varied within the ranges $\beta'_A \in [3 \times 10^{-6},3 \times
10^{-2}]$ and $\gamma'_A \in [2.3 \times 10^{-6},2.3 \times 10^{-5}]$ in the
simulations.  In contrast to the nutrient equations that are solved at the
quasi-stationary state, Eq. $9$ is iterated for a defined number of times, since
the antibodies diffusion constant can be as small as $10^{-7.5}cm^2/s$ in
tissues \cite{Berk}, whereas glucose and oxygen diffusivities are
$10^{-5}cm^2/s$ in tumor spheroids \cite{Nichols}.

\section*{References}

\bibliography{manuscript}

\end{document}